\documentclass[aps,twocolumn,apl,superscriptaddress,showpacs,longbibliography]{revtex4-1}

\usepackage{graphicx}
\usepackage{dcolumn}
\usepackage{bm}
\usepackage{upgreek}
\usepackage{paralist}
\usepackage{gensymb}

\newcommand{\EECS}{\affiliation{Department of Electrical Engineering and Computer Science, Massachusetts Institute of Technology, Cambridge MA 02139}}
\newcommand{\Delft}{\affiliation{Niels Bohr Institute, University of Copenhagen, 2100 Copenhagen, Denmark}}

\begin{document}

\title{Rectangular Photonic Crystal Nanobeam Cavities in Bulk Diamond}
\author{Sara Mouradian}
\email{S.M. and N.H.W. contributed equally to this work}
\author{Noel H. Wan}
\email{S.M. and N.H.W. contributed equally to this work}
\EECS
\author{Tim Schr{\"o}der}
\Delft
\author{Dirk Englund}
\email{englund@mit.edu}
\EECS

\begin{abstract}

We demonstrate the fabrication of photonic crystal nanobeam cavities with rectangular cross section into bulk diamond. In simulation, these cavities have an unloaded quality factor ($Q$) of over $1\times10^6$. Measured cavity resonances show fundamental modes with spectrometer-limited quality factors $\geq 14\times10^3$ within 1\,nm of the NV center's zero phonon line at 637 nm. We find high cavity yield across the full diamond chip with deterministic resonance trends across the fabricated parameter sweeps. 

\end{abstract}

\maketitle

A central aim of quantum information science is the efficient generation of large entangled states of stationary quantum memories with high-fidelity single and two-qubit gates. Among solid-state qubits, a leading system consists of the nitrogen vacancy (NV) center in diamond.  Recently, entanglement between distant NV nodes has been demonstrated via single-photon measurements of the zero phonon line (ZPL) emission~\cite{togan2010quantum,2013Bernien_Nat_entangle}. Such heralded entanglement can be used to build large cluster states which are a resource for universal quantum computation~\cite{barrett2005efficient} or  quantum repeater networks~\cite{childress2005fault,vinay2016practical}. However, the coherent ZPL optical transition of the NV accounts for only ~3\% of the emission due to phonon interactions even at cryogenic temperatures. This severely limits the entanglement rate, even with broadband collection enhancement structures such as a solid immersion lens~\cite{2013Bernien_Nat_entangle}. To improve the entanglement rate, the collection rate into the desired frequency and spatial mode must be increased. Much recent work has focused on maximizing the coherent ZPL collection efficiency via the Purcell effect in a photonic crystal nanocavity.

One major challenge in coupling the NV to a nanocavity is that high-quality diamond cannot currently be grown in thin (wavelength-scale) waveguiding membranes. Hybrid structures have been used to enhance the ZPL emission~\cite{gould2016efficient,englund2010deterministic,wolters2010enhancement}, though the maximum enhancement rate is limited as the NV must be placed out of the cavity's mode maximum. Advances in diamond patterning have enabled the fabrication of photonic crystal cavities in diamond~\cite{Schroder2016_Review} with two main methods:  fabrication into thinned diamond membranes~\cite{faraon2011resonant,faraon2012coupling,hausmann2013coupling,lee2014deterministic,li2015coherent,riedrich2014deterministic,riedrich-moller_one-_2012,li2015nanofabrication} and angled etching of bulk diamond~\cite{bayn2011triangular,2014Igal_APL,burek2014high,schukraft2016imp}. However, the measured quality ($Q$) factors have been limited to a few thousand near the NV ZPL. Low cavity yield or inconsistency across the diamond substrate also presents a major scaling challenges, especially for membrane-based approaches. 

\begin{figure}[ht]
\begin{center}
\includegraphics[width=3.2in]
{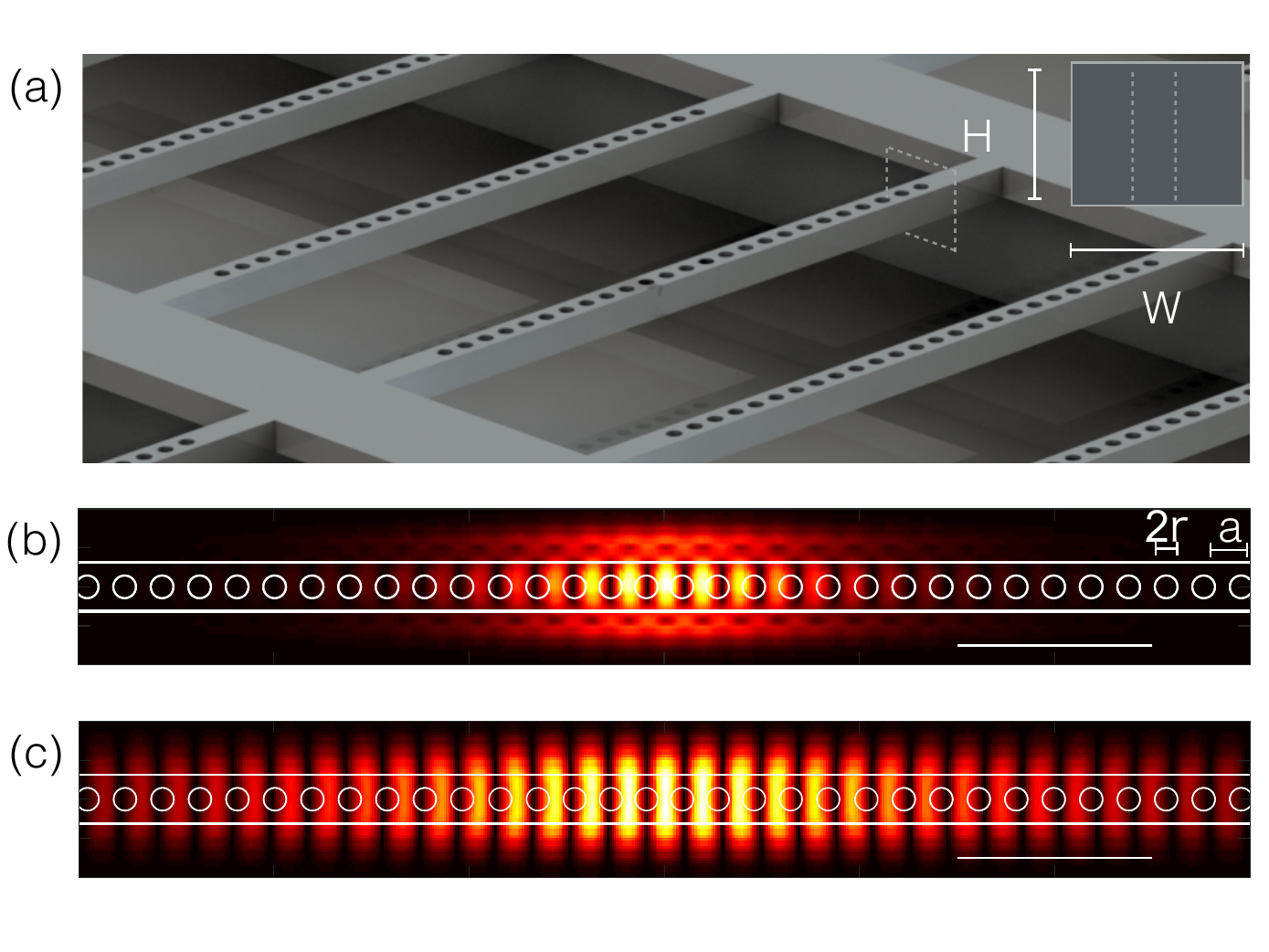}
\end{center}
\caption{(a) Schematic of an array of rectangular nanobeam cavities fabricated from bulk diamond. (b) Re$(E)$ for the first order TE mode. (c) Re$(E)$ for the first order TM mode. Scale bar for (b) and (c): 1 $\upmu$m }
\label{concept}
\end{figure}

Recently, Khanaliloo et al.~\cite{khanaliloo2015high,khanaliloo2015single} introduced a technique to undercut lithographically defined structures in bulk diamond, allowing for the fabrication of high quality free-standing diamond optomechanical devices.  In this Letter, we apply a similar isotropic undercut process to the fabrication of photonic crystal nanobeam cavities. Figure~\ref{concept}(a) shows a schematic of the rectangular nanobeam photonic crystal fabricated from bulk diamond. We find that the quasi-isotropic etching procedure is highly repeatable and consistent across the full chip. In addition, the process requires minimal fabrication optimization as electron-beam lithography determines all parameters except for the nanobeam height, which can be precisely tuned during the relatively slow, isotropic etching. This fabrication process enables instrument-limited optical quality ($Q$) factors exceeding 14,000 within 1\,nm of the NV$^-$ center ZPL wavelength of 637\,nm, as well as uniform nanocavity fabrication across a full chip. 

\begin{figure*}
\begin{center}
\includegraphics[width=6in]
{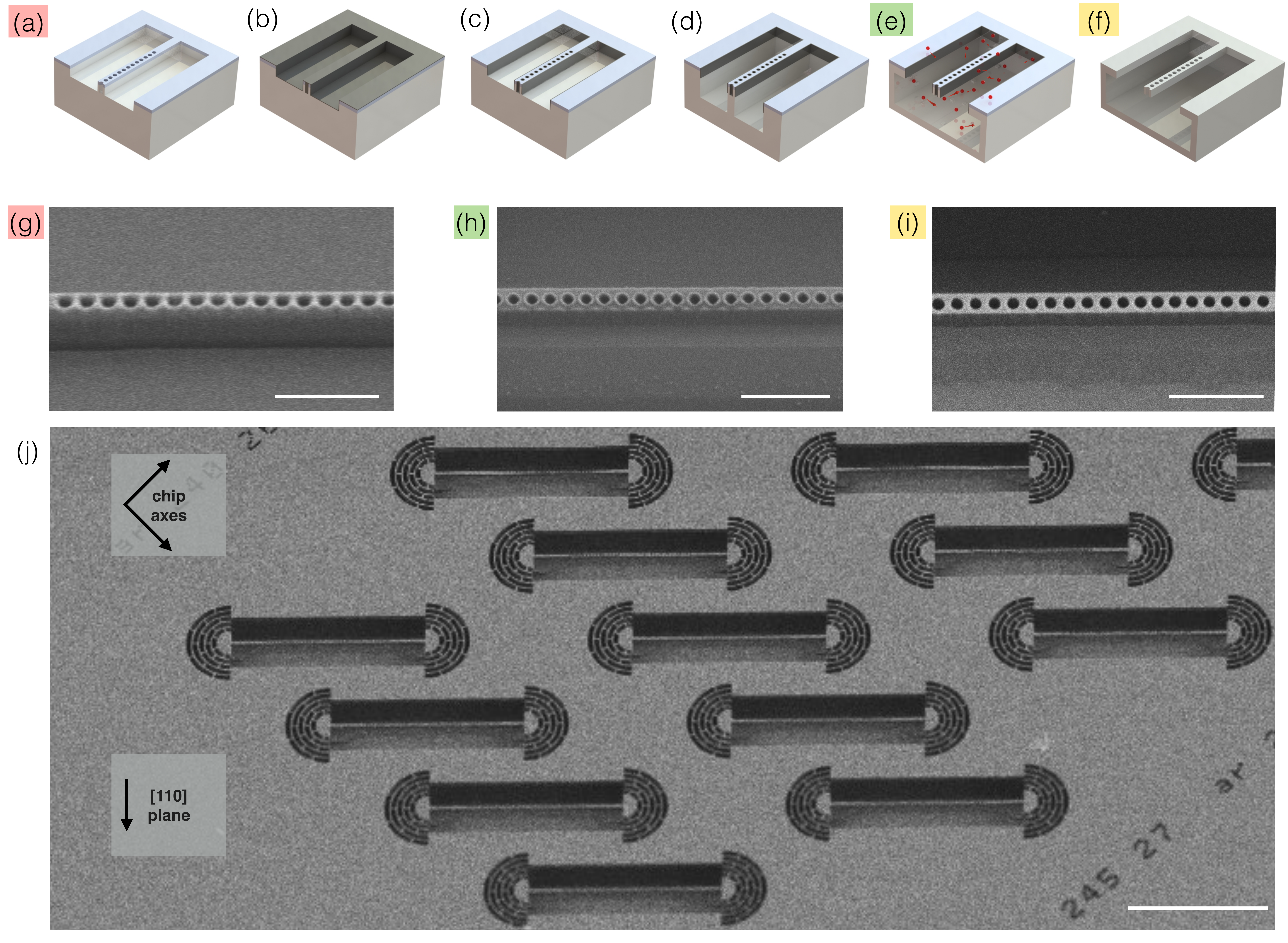}
\end{center}
\caption{(a-f) Fabrication steps for rectangular nanobeam photonic crystal cavities, where white, blue, and gray indicate bulk diamond, SiN, and Al$_2$O$_3$, respectively. (a) Anisotropic etch into diamond with SiN hard mask following electron-beam lithography. (b) Atomic layer deposition of Al$_2$O$_3$. (c) Selective removal of the top layer of Al$_2$O$_3$. (d) Anisotropic etch into diamond (e) Quasi-isotropic etch of diamond nanobeam cavity (f) Suspended nanobeam cavity following mask removal. Scanning electron micrographs (g),(h),(i) correspond to processes (a),(e) and (f), respectively. Scale bar: 1\,$\upmu$m (j) Overview of an array of photonic crystal nanobeam cavities. Scale bar: 10\,$\upmu$m }
\label{fabrication}
\end{figure*}

\begin{figure*}
\begin{center}
\includegraphics[width=6in]{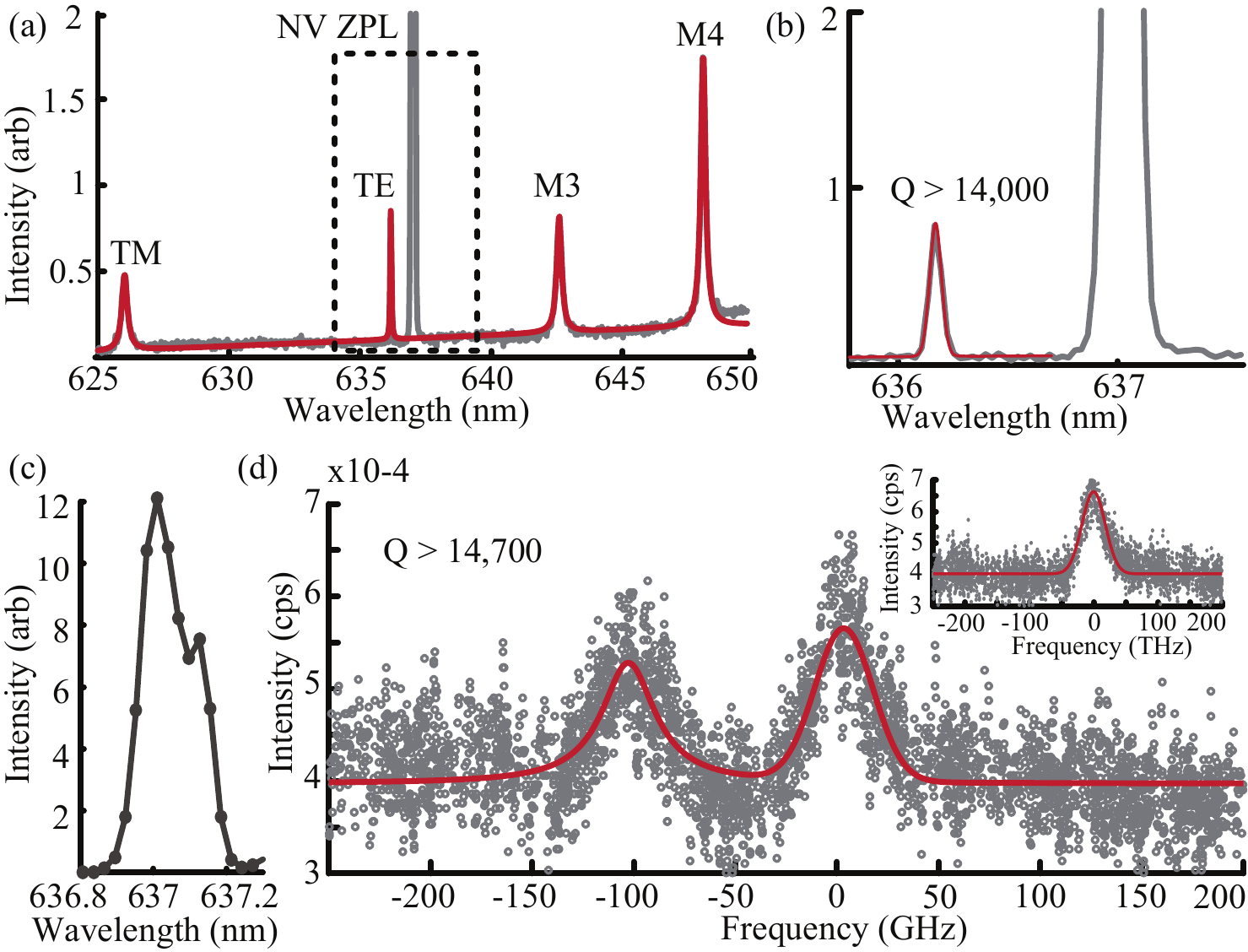}
\end{center}
\caption{(a) PL spectrum of Cavity A with lowest order TE and TM cavity peaks fit with Lorentzian functions. (b) Zoom in of the fundamental TE mode of Cavity A. The red line shows a fit to the spectrometer-limited cavity peak, fit with a pseudo-Voigt function.  (c) PL spectrum of Cavity B showing the inhomogeneously broadened ZPL emission from the ensemble of excited NVs at 637\,nm and a cavity peak at 637.1\,nm. (d) Photoluminescence excitation (PLE) spectrum of Cavity B showing a PLE peak at NV ZPL position and a cavity-enhanced PLE peak 100\,GHz detuned center from the inhomogeneous ZPL distribution. The inset shows a PLE measurement of the same nanobeam 7\,$\upmu$m from the cavity center showing only the inhomogeneous distribution of NV centers in the sample.}
\label{resonances}
\end{figure*}

\begin{figure}[hpt]
\begin{center}
\includegraphics[width=3in]{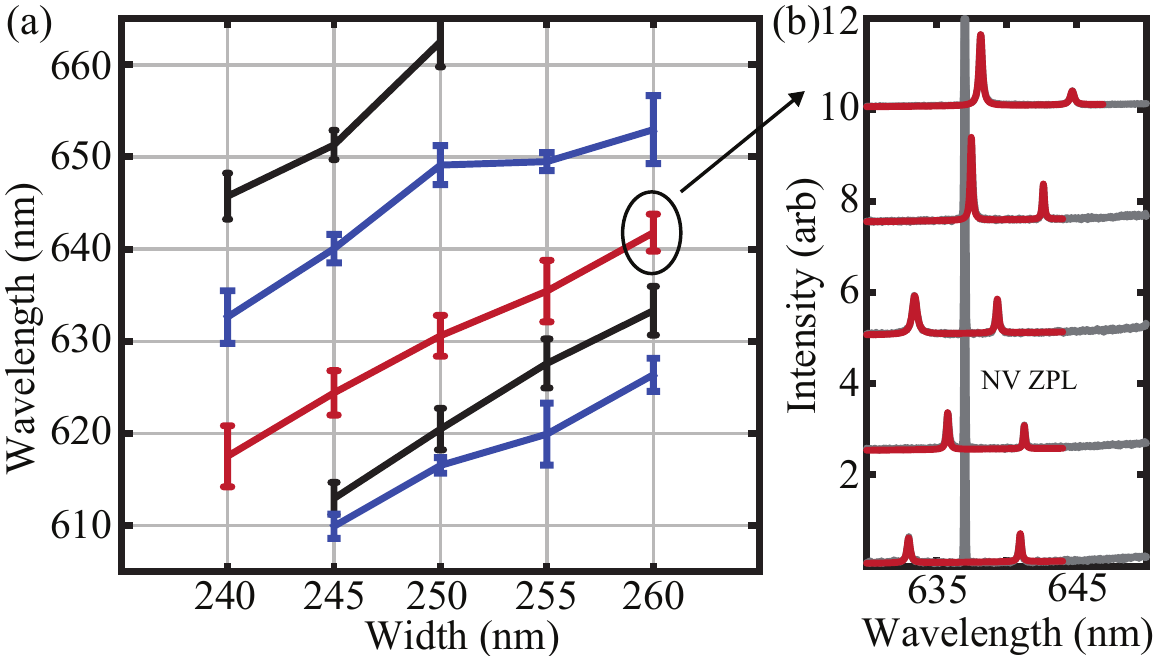}
\end{center}
\caption{(a) Frequency distribution of resonant frequencies of Mode 4 for all 25 sets of parameters. The error bars show the standard deviation $\pm\sigma$ of the resonant frequencies for the 5 cavities fabricated for each set of parameters. (b) The spectra of the 5 cavities fabricated with W = 260\,nm and r = 58\,nm (circled in (a)) showing the distribution of M3 and M4.}
\label{results}
\end{figure}

We design photonic crystal nanobeam cavities to support a low-mode-volume ($V$) mode with high quality factor $Q$ at the NV ZPL (637\,nm), and with the electric-field maximum concentrated in the diamond. The design process begins with approximate cavity parameters derived from band structure simulations and optimizes the cavity $Q$ of the lowest order TE mode (intensity profile shown in Figure~\ref{concept}(b)) at $\lambda_{NV}=\,637$\,nm by Finite Difference Time Domain (FDTD) simulations. The final design consists of a diamond waveguide ($W = 250$\,nm and $H = 230$\,nm) periodically patterned with holes with radius $r = 58$\,nm and spacing $a=192$\,nm. The defect supporting the cavity mode is introduced by linearly decreasing the hole spacing to $a=171$\,nm over 5 periods. With 25 holes on either side of the cavity region, this cavity has a radiation-limited $Q$ factor of $>1$ million. The cavity also supports a TM mode (Figure~\ref{concept}(c) shows the intensity profile) at 615\,nm with a $Q$ of 13,000 in simulation, as well as other TE modes with mode maxima in the outer hole regions with lower $Q$ values -- 5,500 and 2,500 in simulation at 649\,nm and 653\,nm respectively. These higher order modes are denoted as M3 and M4 in the measured spectra in Figures~\ref{resonances} and ~\ref{results}.  

We fabricated these nanobeam cavity designs in a 3\,mm$\times$3\,mm$\times$0.3\,mm single-crystal diamond with a \{100\} top face grown by chemical vapor deposition (CVD, Element6) with a nitrogen defect density of less than 1\,ppm, and thus a native NV density of approximately 1ppb. High-resolution X-ray diffraction confirmed the crystal orientation to align the nanobeams along the facet with the fastest etch rate (\{110\}), which is 45\degree\,from the diamond chip facets on all measured diamond, as illustrated in Fig~\ref{fabrication}(j). The quasi-isotropic etch rate is facet-dependent~\cite{khanaliloo2015single}, and ensures that the beam has a rectangular cross-section. Unlike nanobeam cavities with triangular cross sections which mix the TE and TM modes, these cavities have low out-of-plane scattering loss~\cite{burek2014high,bayn2011triangular}.

Fig.~\ref{fabrication}(a-f) outlines the essential steps of the fabrication process. A 180\,nm-thick low-stress SiN layer, deposited with plasma-enhanced chemical vapor deposition, functions as a hard mask to pattern the diamond. We observed a selectivity of approximately 30:1 for the oxygen anisotropic etch parameters described below. Electron-beam lithography defines the nanobeam cavities (ZEP 520A exposed at 500\,$\mu$C/cm$^2$ and developed at 0$^{\circ}$C in ortho-xylene for 90\,s). Following resist development, a tetrafluoromethane (CF$_4$) plasma reactive-ion etch step transfers the pattern into the the SiN hard mask. This pattern is transferred into the diamond (Figure~\ref{fabrication}(a) using an inductively coupled oxygen plasma (ICP) with a working pressure of 0.15\,Pa with ICP and RF powers of 500\,W and 240\,W, respectively. This anistropic etch step is 2.5 times as deep as the final desired cavity height for precise tuning of the nanobeam height in the subsequent isotropic undercut etch step. This step produces smooth and straight sidewalls as seen in the scanning electron micrograph (SEM) in Fig.~\ref{fabrication}(g).

A conformal layer of 20\,nm of aluminum oxide (Al$_2$O$_3$) produced by atomic layer deposition (ALD), see Fig.~\ref{fabrication}(b), protects all sides of the nanobeam cavity for the subsequent etch steps. A CF$_4$ reactive ion etch removes the top Al$_2$O$_3$ layer, leaving only the sides covered, as shown in Fig.~\ref{fabrication}(c). A second anisotropic oxygen etch, using the same parameters as above, removes an additional 1\,$\upmu$m of diamond, as shown in  Fig.~\ref{fabrication}(d). A quasi-isotropic etch then undercuts the nanobeam structure at 200$^{\circ}$C and 3\,Pa with 900\,W ICP and no forward bias (Fig.~\ref{fabrication}(e)). The elevated temperature and pressure increase the chemical interaction rate with the diamond to increase the etch rate. As seen in the SEM in Fig.~\ref{fabrication}(h), the Al$_2$O$_3$  is thin enough to allow periodic SEM measurements of the nanobeam height. Once the desired nanobeam height of 230\,nm ($\sim$85\,minutes with an intial beam height of 575\,nm) and undercut are achieved, the residual SiN and Al$_2$O$_3$ are removed using 49\% hydrofluoric acid (Fig.~\ref{fabrication}(f)). An SEM of the final structure after mask removal is shown in Fig.~\ref{fabrication}(i). The measured chip contains 125 cavities, with 5 copies at each of 25 parameters. The parameter sweep modified the beam widths and hole radii by $\pm\,4\,\%$ and $\pm\,16\,\%$ respectively, to cover a large wavelength range.

We measured the fabricated cavities at 4\,K by photoluminescence (PL) and photoluminescence excitation (PLE) spectroscopy. The native population of NV centers, excited using a 532\,nm laser, provides an internal light source for the cavity. The cavity PL is collected with a 0.9 NA objective and resolved on a spectrometer with a resolution of 0.06\,nm ($Q \sim$14,000).  Fig.~\ref{resonances} shows two cavity spectra: Cavity A ($W = 260\,$nm, $r = 55\,$nm) and Cavity B ($W = 245$\,nm, $r = 58$\,nm). Fig.~\ref{resonances}(a) shows the cavity-modified PL spectrum of the NV ensemble at Cavity A under 532\,nm excitation. Four modes appear in the spectrum, corresponding to the first-order TE and TM modes (profiles shown in Figure~\ref{concept}(b,c)), as well as the higher-order modes, M3 and M4. Fig.~\ref{resonances}(b) shows the high-$Q$ first-order TE mode at 636.1\,nm, as well as the inhomogeneously-broadened ZPL of the excited ensemble of NV centers at 637\,nm. The linewidth of the cavity mode is limited by the spectrometer's resolution, as confirmed by comparing to the spectrum of a sub-0.5\,MHz laser at the same wavelength. Fitting the spectrum with a pseudo-Voigt function~\cite{albrecht2013coupling} takes into account the effect of the Gaussian spectrometer result on the Lorentzian cavity spectrum. This fit indicates a lower bound of $Q\geq 16700$, above the resolution of the spectrometer ($Q \geq 14,000$). 

The spectrum of Cavity B (Fig.~\ref{resonances}(c)) reveals the first-order TE mode overlapping with the inhomogenous distribution of ZPLs in the NV ensemble. We performed PLE spectroscopy to better measure the $Q$ of this cavity and study the cavity-enhanced excitation of the NV ensemble.  The PLE consisted of NV phonon side band (PSB) detection (filtered $>650$\,nm) while scanning a $< 0.5\,$MHz-linewidth laser over the NV ZPL and cavity frequencies. The inset of Fig.~\ref{resonances}(c) shows the PLE spectrum of the unpatterned nanobeam 7\,$\mu$m from the cavity center. This reveals an inhomogeneously broadened absorption spectrum of the excited population of NV centers centered at 470.48\,THZ (637.2\,nm) with a full width half maximum of 42.7\,GHz. A PLE scan at the center of the cavity shows a second peak that is absent on the rest of the sample; we attribute this to the cavity-enhanced absorption of NV centers coupled to the cavity mode. A Lorentzian fit of the peak reveals a measured cavity $Q$ of 14,700. This provides a lower bound on the $Q$ of the bare cavity mode, as the PLE spectrum may be broadened by the interaction with the inhomogeneous distribution of emitters coupled to this mode of the cavity~\cite{valente2014frequency}.

A survey of all of the fabricated devices demonstrates the consistency and high yield of our fabrication technique. Fig.~\ref{results}(a) shows the measured cavity resonances of mode M4 across the full range of parameters. The error bars show the standard deviation of the wavelength position for the 5 cavities fabricated with each parameter set. We used here mode M4 as it has the highest vertical loss of the four modes considered here, and thus the highest SNR in spectrometer measurements. This survey reveals the expected trends: the resonance wavelengths increase with larger beam widths and decrease with larger holes radii. The standard deviation within each parameter is low with an average of $\pm\,2.2$\,nm deviation from the mean in each parameter set, showing the consistency of the fabrication process. The spectra of M3 and M4 of the 5 cavities with $W=260$\,nm and $r=58$\,nm (circled in Fig.~\ref{results}(a) are shown in Fig.~\ref{results}(b). \\ \\

In summary, we have demonstrated a method to fabricate high-$Q$ photonic crystal nanobeam cavities from bulk diamond at the NV ZPL wavelength. We measured cavity resonances within 1\,nm of the NV ZPL wavelength, with instrument-limited $Q$ factors larger than 14,000. The process showed consistent cavity properties across all fabricated devices. Future work will apply this fabrication process to high-purity diamond with a nitrogen concentration below 100 ppb, which should enable the coupling of individual NV ZPLs with high Purcell enhancement. The rectangular cross section of these cavities should enable efficient mode conversion between diamond waveguides and on-chip ridge or channel waveguides~\cite{mouradian2015scalable}. This fabrication technique applies to other diamond color centers, such as the germanium vacancy and silicon vacancy centers. These emitters have naturally narrow emission linewidth~\cite{schroder2016scalable} so that the nanocavity parameters achieved here should allow for the strong coupling regime of cavity quantum electrodynamics.

\begin{acknowledgments}
This research was supported in part by the Army Research Laboratory Center for Distributed Quantum Information (CDQI). S.M. was supported in part by the NSF IQuISE program and the NSF program ACQUIRE:``Scalable Quantum Communications with Error-Corrected Semiconductor Qubits.'' N.W was supported by CDQI. 
\end{acknowledgments}

\bibliography{2017_CavityFab_v1}

\end{document}